
\def\NP{\vfil\eject}
\def\NI{\noindent}

\magnification=\magstep 2
\overfullrule=0pt
\hfuzz=16pt
\voffset=0.0 true in
\vsize=8.8 true in
\baselineskip 20pt
\parskip 6pt
\hoffset=0.1 true in
\hsize=6.3 true in
\nopagenumbers
\pageno=1
\footline={\hfil -- {\folio} -- \hfil}

\ 

\ 
 
\centerline{{\bf Mechanism of Formation of Monodispersed}}
\centerline{{\bf Colloids by Aggregation of Nanosize Precursors}}
 
\ 

\centerline{Vladimir Privman, Dan V.~Goia, Jongsoon Park and Egon
Matijevi\'c}

\ 
 
\centerline{\sl Center for Advanced Materials Processing}
\centerline{\sl Clarkson University, Potsdam, New York 13699-5814}

\vfill

\NP
 
\centerline{\bf Abstract} 

It has been experimentally established in numerous cases that 
precipitation of monodispersed colloids from homogeneous solutions is a 
complex process. Specifically, it was found that in many systems nuclei, 
produced rapidly in a supersaturated solution, grow to nanosize primary 
particles (singlets), which then coagulate to form much larger final 
colloids in a process dominated by irreversible capture of these 
singlets. This paper describes a kinetic model that explains the 
formation of dispersions of narrow size distribution in such systems. 
Numerical simulations of the kinetic equations, with experimental model 
parameter values, are reported. The model was tested for a system 
involving formation of uniform spherical gold particles by reduction of 
auric chloride in aqueous solutions. The calculated average size, the 
width of the particle size distribution, and the time scale of the 
process, agreed reasonably well with the experimental values.

\NP

\NI{\bf 1.\ \ Introduction}
 
Ever since Faraday reported in 1857 the preparation of gold sols
of different colors (1), scientists have been fascinated by 
monodispersed colloids. However, systematic studies of the synthesis, properties, and mechanisms
of formation of such colloids have been initiated only quarter a century ago. Since then, a large number of uniform dispersions of particles
of simple and mixed chemical compositions and various shapes, ranging in modal size from several nanometers to several micrometers, have been described in the literature. The method of choice has been precipitation from homogeneous solutions, either directly or via gel/sol or sol/gel routes (2-7).

Two essential ingredients are necessary to explain formation of uniform colloids by direct precipitation. The first is to establish the identity of the solute species that are involved in the solid phase formation. The second is to explain all stages leading from the initial nucleation to final particles. Early works on the precipitation of monodispersed colloids were
based on the concept suggested by LaMer, which essentially assumes a short nucleation burst, followed by diffusional growth of the nuclei to form
identical, finely dispersed particles (8,9). This mechanism appeared simple and plausible and was generally accepted as operational.

With the increasing availability of monodispersed colloids of spherical
and other shapes, it has become evident that the 
burst-nucleation/dif\-fusional growth mechanism alone cannot explain 
many experimental observations. The first among these has been the 
finding that some perfectly spherical particles, directly precipitated
from solution, showed X-ray characteristics of known minerals, such as ZnS
of sphalerite (10), CdS of greenockite (11), Fe$_2$O$_3$ of hematite (12), etc. It was obvious that these particles could not be single crystals and, indeed, it has been confirmed by different techniques that a great majority
of monodispersed colloids consist of much smaller crystalline subunits (10-. For instance, small-angle light scattering from CdS particles revealed that they were built of $\sim\,$50$\,$\AA{} uniform globules, while electron microscopy studies substantiated the composite nature of uniform particles such as
CeO$_2$ (13) or SnO$_2$ (14,15). Finally, it has been shown with uniform colloidal copper oxide
that the crystallite size in the final particles, determined by
X-ray diffraction, was the same as the diameter of the precursor subunits (singlets) (16). The latter finding added credence to an aggregation mechanism, rather than to a dissolution/reprecipitation process.

The substructures have been identified also in particles of shapes different from spherical. For example, it was demonstrated that different morphologies of colloidal hematite ($\alpha$-Fe$_2$O$_3$) (12) were related to the natures of the precursor singlets (18-20). 
All this experimental evidence has led to the conclusion that {\it the mechanism of the formation of uniform colloids is more complex than originally expected}, and that it likely involves more than just a nucleation/dif\-fusional growth process. 

Thus, the formation of monodispersed colloids poses two major theoretical challenges. First, the morphology and shape selection of particles formed by interplay of nucleation and aggregation processes must be explained. Second,
the {\it size-selection mechanism}, i.e., the kinetics of generation of narrow particle size distribution, must be identified. Several theoretical approaches utilizing thermodynamic and dynamical growth
mechanisms (4,5,18,21-32) have been described in the literature. 
It has been established that models of aggregation of subunits can be developed that yield a peaked and even sharpening with time particle size distribution (21-34). However, none of these attempts could fit quantitatively a broad range of experimental findings. In this work the first results of a new approach that explains the {\it size selection\/} are reported. Specifically, we argue that the growth by aggregation of subunits must be considered coupled to the process of formation of these subunits, via the time-dependence of the latter.

Theories of coagulation of nanosize particles involve modeling of the particle transport towards each other and particle-particle interactions. The actual attachment event and possible detachment and restructuring processes, which presumably make the density of the final solids close to that of the bulk material, must be then described. There exists extensive recent literature that generally addresses these and related issues in various situations. Let us first mention models of growth at surfaces (35-37) that typically assumed diffusional transport, or convective-diffusional transport, the latter being appropriate for flowing suspensions (38). The main ingredient of the simplest transport models is that the effects of particle-particle interactions are typically incorporated via average, phenomenological Boltzmann weights. Thus, the resulting deposit morphology, which depends on the precise way of particles fitting in and rearranging within the exiting structure, cannot be considered. Instead, the emphasis has been on the fluctuation properties of the topmost, outer growth layers and their scaling description (35-37).

Traditional models of coagulation and nucleation usually assume diffusional transport and consider growth of particles and aggregates either via microscopic nucleation processes (21,39) or by particle-particle, particle-aggregate, and aggregate-aggregate adhesion on encounters (40,41). Models of sparse, percolative aggregates, typically predict the size distribution to peak at small sizes (42). More compact aggregate growth, such as in classical nucleation or coagulation, usually yields size distribution that grows with time, $t$, according to $t^z$. Here $z$ is termed the dynamical exponent and its values range from 0.2 to 0.5, depending on the model. Prescriptions for sharpening the size distribution by actually physically removing particles of certain sizes in the course of the process, in competing nucleation and aggregation, were reviewed in (28).

Recently, the simplest rearrangement processes after attachment, following collision, have been considered, mainly by numerical modeling, including rolling to preferential sites and diffusional relaxation (43-48). However, rearrangement effects on the formation of monosize particles have not been explored. The experimentally observed role of aging in size-selection in fine particle synthesis will be addressed in the following sections. This role is not related to the restructuring of the growing particles.

The rest of this article is organized as follows. In Section~2, a kinetic model of secondary particle formation by singlet-capture dominated growth is considered. In Section~3, primary particle (singlet) formation by burst nucleation as well as other effects are incorporated to complete the model. Section~4 describes the gold sol used as the experimental system to test the model. Finally, Section~5 reports numerical simulations and comparisons with the experiments.
 
\NP 

\NI{\bf 2.\ \ Kinetic Model of Particle Formation}
 
Uniform particle formation in colloid synthesis usually proceeds in several stages. In the initial induction stage solutes are formed to yield a supersaturated solution, leading to nucleation. The nuclei then grow by diffusive mechanism to form the primary particles (singlets), which in turn aggregate to form secondary particles. This latter process is sometimes facilitated by changes in the chemical conditions in the system. For example, the ionic strength may increase, or the pH may change, causing the surface potential to approach the isoelectric point. Both effects
result in a reduction of electrostatic barriers, thus promoting particle aggregation. Formation of the final (secondary) particles, which can be of narrow size distribution, is clearly a diffusion-controlled process (2-7). Some of the various stages of the process can be more appropriately viewed as {\it particle\/} evolution rather system's dynamics.

An important assumption of the present model is the experimentally observed property that the secondary particles are sufficiently sparsely positioned in solution to conclude that their evolution is largely independent. Furthermore, it is assumed that the particles are spherical, with density close to that of the bulk material, which is experimentally a very common case, and the focus will be on their size distribution due to growth
by aggregation. These approximations will be discussed at the end of this section. 
Several other simplifying assumptions are made
in order to zero-in on the essential ingredients of the theoretical modeling and make numerical simulations tractable. For instance, it is possible that singlets forming secondary particles may undergo restructuring and rearrangements resulting in more compact structures and leading to shape selection. Such processes are neglected in the present model, but may be incorporated in  future studies.

Thus, let us focus on the two main stages in the process: the production of primary and secondary particles. The latter will be treated presently, while the formation of nuclei and their growth (aging) to 
primary particles will be considered in the next section.
Kinetic rate-equation-type models of aggregation typically utilize a master equation for the distribution of growing particles by their size. In the present model, the size will be defined by how many primary particles (singlets) were aggregated into each secondary particle, denoted by $s=1,2,\ldots$. The growing particles can adsorb or emit singlets and multiplets. The master equation is then quite standard to set up, as in, e.g., (33). In the present case, the process is experimentally documented to be highly irreversible, so that detachment can be disregarded. 

Furthermore, it is assumed that the singlets have diffusion constants larger than aggregates so that their attachment dominates the process; 
this approximation will be discussed later. The master equation then takes the form

$$ { d  N_s \over  d  t } = w_{s-1} N_{s-1} - w_s N_s \; , \quad \quad {\rm for\ \ } s > 1 \; . \eqno[1] $$

\NI Here $N_s (t)$ is the time-dependent number density (per unit volume) of the secondary particles consisting of
$s$ primary particles. The attachment rate $w_s$ will be taken from the Smoluchowski expression (49):

$$ w_s (t) = 4 \pi R_s D N_1 (t) \; , \eqno[2] $$

\NI where $R_s$ is the radius of the $s$-size particle, given by

$$ R_s = 1.2\, r s^{1/3} \; . \eqno[3] $$

\NI The parameters $r$, the primary particle radius, and $D$, their diffusion constant, are experimentally available. The constant 1.2
in Eq.~[3] was calculated as

$$(0.58)^{-1/3}\simeq 1.2 \; , \eqno[4] $$

\NI where 0.58 is the typical filling factor
of the random loose packing of spheres (50). The fact that all primary particles were assumed to have the same radius $r$ will be 
addressed in Section~3. 
Note that the rate in Eq.~[2] involves the product $R_s D \propto rD$, where the latter, according to the Einstein formula for the diffusion constant, is not sensitive to the distribution of values of the radii $r$.

In standard approaches to aggregation, the evolution of the population of singlets, which is
not covered by Eq.~[1], is obtained from the conservation of matter:

$$ N_1 (t) + \sum\limits_{j=2}^\infty j N_j(t) = N_1 (0) \; , \eqno[5] $$

\NI which assumes that initially, at time $t=0$, there are only singlets. In this approach, however, singlet-dominance cannot be assumed; specifically, Eq.~[1] and/or Eq.~[5] need to be modified. Otherwise, the particle distribution is confined to small sizes, as
confirmed by numerical simulations and other studies (46). Indeed, most singlets will simply combine into dimers, fewer trimers, etc., then the growth stops. 
The conventional approach has been to consider more general models, with discrete or continuous population balance, allowing multiplet-multiplet
aggregation, which adds terms in Eq.~[1]; see, e.g., (21-34). These models will be commented on at the end of this section. Frequently, though, multiplet-multiplet aggregation models yield wide
particle-size distributions, of the scaling type (33,35-37), and they cannot explain the size selection.

In this work 
we develop a different approach based on the observation that the supply of ``singlets,'' i.e., of primary particles, is in itself a dynamic process. Numerical calculations indicate that if the concentration of the singlets were constant, i.e., if they were continuously generated to compensate for their depletion due to aggregation, then the resulting particle size distribution would be wide and peaked at small sizes, such that $N_s \simeq N_1 s^{-1/3}$ up to some growing cut-off $s$-value.
However, {\it if the availability of singlets is controlled to be a decaying function of time, then size selection can be obtained in some cases}. For instance, when the rate $\rho (t)$ at which the primary particles are formed (per unit volume) was chosen to be decaying according to a power-law, then
preliminary numerical calculations yielded a single-hump size distribution for the secondary particles. Note that the equation for $N_1(t)$ must be modified by replacing
Eq.~[5] with the relation

$$ N_1 (t) = \int\limits_0^t \rho (t^\prime)\, d t^\prime - \sum\limits_{j=2}^\infty j N_j(t) \; , \eqno[6] $$

\NI with the initial values $N_s(0)=0$ for all $s=1,2,3,\ldots$. 

Equations of the type considered here, with singlet-capture dominance of the dynamics 
and several ad hoc singlet-input rate functions, have been described in the literature (51). The emphasis of these studies has been on cases which are exactly solvable, for instance, one-dimensional versions, and those where the size distribution shows self-similar ``scaling'' behavior.
The goal of the present work is quite different, i.e., to actually {\it model\/} the primary-particle input rate $\rho(t)$ by using the burst-nucleation approach; see the next section.

Let us now address some of the approximations involved in our treatment of the aggregation process. These include, for instance, ignoring multiplet mobility and multiplet-miltiplet collisions,
as well as the effect of mobility of aggregates (multiplets) on the diffusion constants used in the rate expressions, etc., especially in 
the beginning of the process when most aggregates are small. There are established methods in the literature that avoid many of such difficulties (21-34) and, in fact, some of the models for various reaction rates lead to particle-size distributions peaked and even sharpening with time. 
However, the main point of our 
work, which we believe is new and crucial to obtaining narrow size
distributions, has been that the growth of the
secondary particles must by coupled, via the rate of generation of
singlets (primary particles), to the time-dependence of the process of 
formation of the latter; see the following sections.

Thus, we intentionally took the {\it simplest possible\/} models of both processes, the primary and secondary particle formation. This choice has the following advantages: it simplifies numerical simulations and thus 
allows to scan a wider range of model parameters; it avoids introduction
of unknown microscopic parameters. As a result, for instance, our description
of the aggregation process in this section has {\it no adjustable parameters};
they are all experimentally available. 

Similarly in the following sections, for primary particle formation, we only utilize {\it one adjustable parameter},
the effective surface tension, and even that parameter turns out to be
close to the experimental bulk value.
However, we recognize that more sophisticated modeling can
improve consistency with experiment, 
perhaps at the expense of additional assumptions and parameters, and we 
intend to explore this avenue
of investigation in future studies. In fact, we carried out preliminary numerical simulations allowing for the dimer diffusion and attachment to larger
aggregates and showing trend of improved consistency with the experiment.

We furthermore note that the approximation of restricting the aggregation
process to only the smallest particles sticking to the larger particles 
has been widely used in the literature, for instance, (22,27). 
Considerations of colloid stability have been typically used to justify
such approximations, and we note that detailed arguments of this sort would require additional microscopic parameters in the model.

We also comment that generally in diffusion-limited growth the aggregates
are expected to be fractal (35-37,52). In this work we avoid
the issue of shape selection and internal structure of the secondary
particles; see a review (21). There is obviously some restructuring going
on during the secondary particle growth for the experimental system
described in Section~4, because the particles are clearly spherical throughout the growth process. Actually, in some other experiments, e.g., (14), the surface of the secondary particles is initially ``hairy'' and
it gets smoother at later times. In our case the internal restructuring processes are fast enough so that the shape/morphology of the aggregates is always spherical/compact. One could propose that  
the effects of restructuring in our model make the 
coefficient in Eq. [3] an adjustable parameter rather than a fixed number. We did not explore this matter.

\NP 

\NI{\bf 3.\ \ Modeling the Primary Particle Formation}

Primary-particle production rate $\rho (t)$ will be calculated by using the relations available in the literature (8,9,34,53) for the fast nucleation process. Such thermodynamic models of ``burst'' nucleation of more or less uniform dispersions have been originally formulated in (8,9). Modeling the rate of formation of primary particles (singlets) requires setting up a master equation, where the rate of growth is determined by the Boltzmann factor with the thermodynamic free energy difference $\Delta G$, multiplied by $-1/(kT)$, in the exponent. This approach in turn requires modeling of the free energy of the growing embryos (sub-critical nuclei); in the simplest model one can use generic volume-plus-surface energy forms.

Let us term ``solutes'' the species (atoms, ions, molecules) which serve as monomers for the primary-particle nucleation. Given the concentration $c(t)$ of solutes, which is larger than the equilibrium ``saturation'' concentration $c_0$ and approaches $c_0$ for large times $t$, the rate of formation
of critical nuclei can be written as (34,53)

$$ \rho(t)=4\pi a n_{cns}^{1/3} {\cal D} c^2 e^{-\Delta G_{cns} /kT} \; , \eqno[7] $$

\NI which is based on the diffusional capture of solutes, whose effective radius is denoted by $a$, diffusion constant by ${\cal D}$, and $n$ is the number of solutes in an embryo. The subscript $cns$ refers to values calculated at the {\bf c\/}ritical {\bf n\/}ucleus {\bf s\/}ize. Note that $c(t)=n_1(t)$. 

The expression~[7] involves the following ingredients. For embryos of size $n < n_{cns}$, one assumes that solutes can be captured and emitted fast enough so that the
size distribution is given by the equilibrium form. Thus,
the factor $ce^{-\Delta G_{cns} /kT}$ in Eq.~[7] follows
from the assumption that sizes up to $n_{cns}$ are ``thermodynamically'' distributed, according the Boltzmann form. For sizes larger than $n_{cns}$ the dynamics is usually assumed to be fully irreversible, corresponding to an unbound growth by the capture of solutes.
The factor $4\pi a n_{cns}^{1/3} {\cal D} c$ in Eq.~[7] is thus the appropriate version of the Smoluchowski growth rate similar to that in Eqs.~[2,3]. The filling-fraction correction factor was absorbed in the definition of $a$ to simplify the notation; it will be specified in Section~5.

For the free energy of the $n$-solute embryo, the known expression
is used:

$$ \Delta G = - n kT \ln \left( c / c_0 \right) + 4 \pi a^2 n^{2/3} \sigma \; , \eqno[8] $$

\NI which involves the bulk term, proportional to $n$, and the surface term.
The standard form of the bulk term was derived as follows. It is assumed that the entropic part of the free-energy change between the solid and solution phases can be calculated as the entropy in the supersaturated liquid suspension of solutes of concentration $c$, as given by the dilute (noninteracting) expression of the ``entropy of mixing,'' defined, e.g., in (54). The surface term in Eq.~[8] corresponds to the assumption that the growing embryos are spherical, of radius $a n^{1/3}$, and introduces their effective surface
tension $\sigma$, which is usually assumed to be comparable to the bulk surface tension.
It will be shown later that in the present case the results are very sensitive to the value of $\sigma$. 
Obviously, all the above expressions only apply for large $n$. It has been a common practice in the literature to use them for all $n$, as one of the approximations involved in a model.

It follows that both  $n_{cns}$ and $\Delta G_{cns}$
are {\it explicit functions\/} of $c(t)$, 

$$ n_{cns}=\left[ 8 \pi a^2 \sigma \over 3 kT \ln \left(c/c_0\right) \right]^3 \; , \eqno[9] $$

$$ \Delta G_{cns} = {256 \pi^3 a^6 \sigma^3 \over 27 (kT)^2 \left[\ln \left(c/c_0\right) \right]^2}
\; , \eqno[10] $$

\NI where the critical value $n_{cns}$ was calculated from $ \partial G/\partial n=0$.

As the next step, the decrease in the concentration of solutes owing to the formation
of critical nuclei is taken into account. Usually, one accepts that for $n > n_{cns}$ the primary particles grow (age) largely by absorbing diffusing solutes, 
and as in the preceding section we ignore here more complicated processes such as capture of small embryos, dissolution, etc. Simultaneously, the primary particles are also captured by the
secondary particles. In the present model, it is assumed for simplicity that the primary particles are captured fast enough by the growing secondary particles so that the effect of their aging on the concentration of solutes can be ignored. Furthermore, it has been generally recognized that aging, when significant, tends to sharpen the size distribution (21,34). Thus, the primary particle radius $r$, introduced in the preceding section, will be assumed to have a single, experimentally determined value, although in reality they have a finite-width, albeit narrow, size distribution. These are among several simplifications on which the present model is based; they can be relaxed in later studies.

Accordingly, one can write

$$ {dc \over dt}=-n_{cns} \rho(t) \; , \eqno[11] $$

\NI which means that the concentration of solutes is ``lost'' solely due to the irreversible formation of the critical-size nuclei. Collecting all the above expressions, one gets the following equations for $c(t)$ and $\rho(t)$:

$$ {dc \over dt}= - {16384 \pi^5 a^9 \sigma^4 {\cal D} c^2 \over 81 (kT)^4 \left[\ln \left(c/c_0\right) \right]^4}
\exp \left\{ -{256 \pi^3 a^6 \sigma^3 \over 27 (kT)^3 \left[\ln \left(c/c_0\right) \right]^2}
\right\} \; , \eqno[12] $$

$$ \rho (t) = {32 \pi^2 a^3 \sigma {\cal D} c^2 \over 3 kT \ln \left(c/c_0\right) }
\exp \left\{ -{256 \pi^3 a^6 \sigma^3 \over 27 (kT)^3 \left[\ln \left(c/c_0\right) \right]^2}
\right\} \; . \eqno[13] $$

It should be noted that replacing the distribution of the primary particle sizes by a single, experimentally measured average value of $r$, violates the conservation of matter. Thus, {\it in the present model only the shape of the secondary particle size distribution is relevant}. The absolute number densities $N_s$ must be rescaled to correspond to the actual amount of the solid matter per unit volume. The latter data are usually available experimentally.
 
\NP 

\NI{\bf 4.\ \ Experimental System}

In order to test the model developed in Sections 2 and 3, dispersions of spherical 
gold particles were used, which were produced by the reduction of chloroauric acid 
(HAuCl$_4$) with ascorbic acid (55).  The simplicity of the chemical reactions involved, the 
stability of the solid phase formed, and the possibility to either measure or estimate all the 
necessary parameters make this system suitable for testing the described 
model.

Several methods of preparation of stable monodispersed gold sols by the 
reduction process have been reported (56-58). Most of these studies have been focused on 
approaches that avoided the aggregation process (i.e., very dilute and/or heavily stabilized 
systems). In contrast, in concentrated dispersions, the aggregation process is assured, 
resulting in the formation of particles consisting of a large number of subunits. The 
precipitation procedure used in the described study (55) has resulted in large spherical 
gold particles of a narrow size distribution. After a short induction period
(up to 6-8 sec) the nucleation occurs, followed by immediate aggregation. The total reaction time varies from 3 to 20 sec, depending on the experimental conditions selected.
Scanning electron micrographs in 
Figure 1 illustrate the resulting particles, at two different magnifications, obtained by rapid 
addition of chloroauric acid to a solution of ascorbic acid under intense agitation.

The field emission microscopy, Figure 2, clearly reveals the presence of the 
subunits, having an approximate size of 30 to 40 nm. The 
calculated value for the packing fraction of the subunits in the 
aggregated particles, $\sim 58$\%, was based (50) on the experimental tapped density of dry powder. This value is characteristic of a random loose packing, 
usually expected in the formation of rapidly assembled systems (50).  The size of the 
primary particles can be also estimated from the X-ray diffraction measurements by the Scherrer 
formula, if one assumes that the subunits are perfect crystals.  Several calculations, using 
different peaks, have generated values between 30 and 42 nm, which is in excellent 
agreement with the data from electron microscopy.  Finally, in an attempt to prepare and 
characterize the primary gold particles before undergoing aggregation, the same 
precipitation process was carried out in very dilute solutions where the metal particles 
were electrostatically stabilized. Figure~3 shows 
the size distribution of these particles, as determined by dynamic light scattering.  Again, 
the modal diameter of $\sim 40\,$nm agrees with the previous values for the precursor 
subunits.  

\NP

\NI{\bf 5.\ \ Results and Discussion}

Numerical simulations
required to follow the time evolution of the kinetic equations turned out to be large-scale. Therefore, the present testing of the model has been restricted to one randomly selected set of experimental parameters. 
However, we actually varied numerically all the
parameter values and found that the calculation results are affected to various degree by them. The parameter values will be discussed roughly in the order of increasing sensitivity of the numerical results.
For consistency all input data are in the MKS system of units.

The radius of the primary particles,

$$ r=2.10 \cdot 10^{-8} \, {\rm m} \; , \eqno[14] $$

\NI was obtained as described in the preceding section, which is
within the range of $0.5\cdot 10^{-8} \, {\rm m}$ to
$5\cdot 10^{-8} \, {\rm m}$, typical for the system under consideration. 
The value in Eq.~[14] applies to the experiment for which the initial concentration $c(0)$ was

$$ c(0)=6.0 \cdot 10^{25} \, {\rm m}^{-3} \; ; \eqno[15] $$

\NI $ c(0)$
was calculated from the concentration of the gold solution used in the preparation of the dispersion, and it yields the initial condition in Eqs.~[12,13]. The diffusion constant $D$ of the primary particles was obtained from the Einstein formula,

$$ D=1.03 \cdot 10^{-11} \, {\rm m}^2\, {\rm sec}^{-1} \; , \eqno[16] $$

\NI with

$$ kT=4.04 \cdot 10^{-21} \, {\rm J} \; . \eqno[17] $$

The saturation concentration of gold in solution, $c_0$, is not well known and is expected to depend somewhat on the experimental conditions.
Using $2\cdot 10^{-12}\,$mol$\,$dm$^{-3}$ \  (59) yields 

$$c_0 \simeq 1\cdot 10^{15} \, {\rm m}^{-3} \; . \eqno[18] $$

\NI The results for the particle size distribution are not particularly sensitive to this parameter, because it enters under logarithm in Eqs.~[12,13]. The solute diffusion constant, 

$$ {\cal D}=1.5 \cdot 10^{-9} \, {\rm m}^2 \, {\rm sec}^{-1} \; , \eqno[19] $$

\NI was estimated similarly to $D$ in Eq.~[16], using the
Einstein formula, with the radius
of the gold atom of $1.44 \cdot 10^{-10} \,$m (60,61).
The applicability of this formula to single atoms may not be exact, but the particle size distribution is not too sensitive
to this parameter value: a decrease in ${\cal D}$ shifts the calculated distribution to somewhat smaller aggregate sizes.

It was established numerically that the size distribution of the secondary particles was sensitive to the values of $a$, the effective atomic (solute) radius,
and to the surface tension $\sigma$.
Note that $a$ was defined to relate
the number of solutes $n$ in a growing primary particle to its radius, given by $an^{1/3}$. It is assumed that the primary particles are largely crystalline; thus, the best choice of $a$ is such that $4\pi a^3
/3$ is the volume per atom, including the attributable part of the surrounding void volume, in bulk gold. Consequently,

$$ a=1.59 \cdot 10^{-10} \, {\rm m} \;  \eqno[20] $$

\NI was obtained by dividing the radius of the gold atom
($1.44 \cdot 10^{-10} \, {\rm m}$) by the cubic root of the volume filling fraction, 0.74, of the crystalline structure of gold (50). 

The effective surface tension of nanosize gold embryos in solution, $\sigma$, profoundly affects the numerical results. Unfortunately, even the bulk-gold value, which is of order

$$ \sigma \simeq 0.58 \; {\rm to}\; 1.02 \, {\rm N}\,{\rm m}^{-1} \; , \eqno[21] $$

\NI is not well known (62), and it may differ from that of the ``nanosize'' solids. Given
this fact, $\sigma$ was chosen as the only adjustable
parameter in the model.

Experimentally, the time scale on which the secondary particle growth effectively terminated was about 8 to $10 \, {\rm sec}$, which does not include the ``induction'' stage. In Figures 4 though 6, the results of the numerical simulations of the kinetic equations are presented with parameters as specified above, for three different values
of $\sigma$, which clearly demonstrate the sensitivity to the choice of this parameter. 
In Figure 4, the case $\sigma=0.51\,$N$/$m illustrates growth that already reached saturation
for times up to $10 \, {\rm sec}$. It should be noted that the distribution evolves quite slowly with time. Initially, it is heavily weighed in the small-aggregate regime. Later on, the large-size peak develops and eventually dominates the distribution.
By varying $\sigma$ near the expected range, given in Eq.~[21], it was found that, for times up to $10 \, {\rm sec}$, {\it all\/} $\sigma$ values yielded smaller average sizes than the experimentally measured one,

$$ \left( R_s \right)_{\rm average\ (experimental)} = 1.0 \pm 0.1 \, \mu{\rm m} \; . \eqno[22] $$

\NI Seeking $\sigma$ that would give the largest secondary-particle size resulted in 

$$ \left( \sigma \right)_{\rm fitted} =  0.57 \pm 0.04 \, {\rm N}\,{\rm m}
^{-1} \; , \eqno[23] $$

\NI which agrees well with the bulk value in Eq.~[21]. 

Figure 5 shows the size distribution for $\sigma=0.57\,$N$/$m. The growth did not reach full saturation on the relevant time scales, and
the peak particle radius at $t=10 \, {\rm sec}$, of $R_s \simeq 0.32 \, \mu{\rm m}$, is somewhat smaller than the experimental value in Eq.~[22]. The width of the distribution, of $\sim$10$\,$\%{}, is close to that established experimentally. Considering the approximations involved in the model, only semiquantitative agreement with the experimental data should be expected. Since the key feature of the model is the prediction of the narrow-width distribution of secondary particle sizes, the overall consistency with the experimental results is gratifying.

As the value of $\sigma$ is increased, the large-size peak does not fully develop on the relevant time scales, as exemplified by the case $\sigma=0.63\,$N$/$m shown in Figure 6. The reader should be reminded that, owing to the absence of the conservation of matter in this model, the number densities $N_s$ must be rescaled according to the actual amount of the solid matter per unit volume, if the comparison of the calculated and experimental distribution is attempted. This refinement will be addressed in future studies.

Asymptotically, the particle-size distribution
``freezes'' for large times, i.e., the particle growth actually stops in this model as opposed to the ``scaling'' growth studied, for instance, in (51). To demonstrate this property, note that Eq.~[11], with Eq.~[9], can be integrated in closed form to yield

$$ \int\limits_0^t \rho (t^\prime) \, d t^\prime = \left( { 3kT
\over 8 \pi a^2 \sigma } \right)^3 c_0 \left[ F\left( {c(0)
\over c_0 } \right) - F\left( {c(t)
\over c_0 } \right) \right] \; , \eqno[24] $$

\NI where 

$$ F(x)=x\left[ \left( \ln x \right)^3 - 3 \left( \ln x \right)^2
+6 \left( \ln x \right) -6 \right] \; . \eqno[25] $$

\NI The left-hand side of Eq.~[24] is the total number of primary particles produced by the time $t$. Obviously, from Eqs.~[24,25] this number is finite as $t \to \infty$, i.e., in the limit $c(t) \to c_0$. The supply of the primary particles, manifested by the peak at small sizes for short times,
in Figures 4-6, is essential for the large-size peak in the distribution to develop and grow, because the present model assumes the growth of the secondary particles to be solely by singlet capture.

An interesting mathematical construction is suggested by ideas developed in
(51). Let us consider the quantity $\tau$ defined by

$$ \tau = \int\limits_0^t 
{N}_1(t^\prime)\, dt^\prime \; . \eqno[26] $$

\NI If the independent variable is changed from $t$ to $\tau$
in Eq.~[1-4], the resulting equations are linear in $N_{s>1}$. One can then easily show that the only way to have a normalizable stationary size-distribution as $t \to \infty$
is to have $\tau (t)$ approach a {\it finite\/} value for large $t$. This quantity was calculated numerically for the $\sigma$ values used in Figures
4-6, and the results, shown in Figure~7, confirm the earlier observations. Specifically, the growth process saturates fast for $\sigma=0.51\,$N$/$m. For the two larger $\sigma$ values there is still some variation for the time scales of order 1 to 10$\,$sec. This function is useful in identifying the time scales of the growth process. 

In summary, a model is proposed for the synthesis of particles with the size distribution that is peaked at an average value corresponding to a large number of primary particles in a final secondary particle. For the experimental gold-sol system the model has worked reasonably well: the average size, the width of the distribution, the time scale of the process, and even the fitted effective surface tension were all semiquantitatively consistent with the measured, known, or expected values. The present model has involved numerous simplifying assumptions. Further studies are needed to incorporate additional effects in the model and test it for a wide range of experimental systems. The main conclusion has been that multistage growth models can yield size-selection as a kinetic phenomenon, which has been observed in a large number of experimental systems.

\NP

\centerline{\bf References}{\frenchspacing

\NI\hang 1. Faraday, M., 
{\it Philos. Trans. R. Soc. London, Ser. A\/} {\bf 147}, 145 (1857).

\NI\hang 2. Matijevi{\' c}, E., 
{\it Ann. Rev. Mater. Sci.\/} {\bf 15}, 483 (1985).

\NI\hang 3. Haruta, M. and Delmon, B., 
{\it J. Chim. Phys.\/} {\bf 83}, 859 (1986).

\NI\hang 4. Sugimoto, T., 
{\it Adv. Colloid Interface Sci.\/} {\bf 28}, 65 (1987).

\NI\hang 5. Sugimoto, T., 
{\it J. Colloid Interface Sci.\/} {\bf 150}, 208 (1992).

\NI\hang 6. Matijevi{\' c}, E., 
{\it Chem. Mater.\/} {\bf 5}, 412 (1993).

\NI\hang 7. Matijevi{\' c}, E., 
{\it Langmuir\/} {\bf 10}, 8 (1994).
 
\NI\hang 8. LaMer, V. K., 
{\it Ind. Eng. Chem.\/} {\bf 44}, 1270 (1952).

\NI\hang 9. LaMer, V. K. and Dinegar, R. H., 
{\it J. Amer. Chem. Soc.\/} {\bf 72}, 4847 (1950).

\NI\hang 10. Murphy-Wilhelmy, D. and Matijevi{\' c}, E., 
{\it J. Chem. Soc., Faraday Trans. I\/} {\bf 80}, 563 (1984).

\NI\hang 11. Matijevi{\' c}, E. and Murphy-Wilhelmy, D., 
{\it J. Colloid Interface Sci.\/} {\bf 86}, 476 (1982).

\NI\hang 12. Matijevi{\' c}, E. and Scheiner, P., 
{\it J. Colloid Interface Sci.\/} {\bf 63}, 509 (1978).

\NI\hang 13. Hsu, U. P., R\"onnquist, L. and Matijevi{\' c}, E., 
{\it Langmuir\/} {\bf 4}, 31 (1988).
 
\NI\hang 14. Oca\~na, M. and Matijevi{\' c}, E., 
{\it J. Mater. Res.\/} {\bf 5}, 1083 (1990).

\NI\hang 15. Oca\~na, M., Serna, C. J. and Matijevi{\' c}, E., 
{\it Colloid Polymer Sci.\/} {\bf 273}, 681 (1995).

\NI\hang 16. Lee, S.-H., Her, Y.-S. and Matijevi{\' c}, E.,
{\it J. Colloid Interface Sci.\/} {\bf 186}, 193 (1997).

\NI\hang 17. Edelson, L. H. and Glaeser, A. M.,
{\it J. Am. Chem. Soc.\/} {\bf 71}, 225 (1988).

\NI\hang 18. Bailey, J. K., Brinker, C. J. and Mecartney, M. L.,
{\it J. Colloid Interface Sci.\/} {\bf 157}, 1 (1993).

\NI\hang 19. Morales, M. P., Gonz\'ales-Carre\~no, T. and Serna, C. J., 
{\it J. Mater. Res.\/} {\bf 7}, 2538 (1992).

\NI\hang 20. Oca\~na, M., Morales, M. P. and Serna, C. J.,
{\it J. Colloid Interface Sci.\/} {\bf 171}, 85 (1995).

\NI\hang 21. Dirksen, J. A. and Ring,  T. A.,
{\it Chem. Eng. Sci.\/} {\bf 46}, 2389 (1991).

\NI\hang 22. Dirksen, J. A., Benjelloun, S. and Ring, T. A.,
{\it Colloid Polym. Sci.} {\bf 268}, 864 (1990).

\NI\hang 23. Ring, T. A., {Powder Technology\/} {\bf 65}, 195 (1991).

\NI\hang 24. van Blaaderen, A., van Geest, J. and Vrij, A.,
{\it J. Colloid Interface Sci.\/} {\bf 154}, 481 (1992).

\NI\hang 25. Bogush, G. H. and Zukoski, C. F., 
{\it J. Colloid Interface Sci.\/} {\bf 142}, 1 (1991).

\NI\hang 26. Bogush, G. H. and Zukoski, C. F., 
{\it J. Colloid Interface Sci.\/} {\bf 142}, 19 (1991).

\NI\hang 27. Look, J.-L., Bogush, G. H. and Zukoski, C. F.,
{\it Faraday Discuss. Chem. Soc.\/} {\bf 90}, 345, and {\it General
Discussion\/} section, 377-383 (1990).

\NI\hang 28. Randolph, A. D. and Larson, M. A., ``Theory of 
Particulate Processes,'' Academic Press, San Diego, 1988.

\NI\hang 29. Brinker, C. J. and Scherer, G. W., ''Sol-Gel Science,''
Academic Press, Boston, 1990.

\NI\hang 30. Flagan, R. C., {\it Ceramic Trans.\/} {\bf 1(A)}, 229
(1988).

\NI\hang 31. Scott, W. T., {\it J. Atmospheric Sci.\/} {\bf 25},
54 (1968).

\NI\hang 32. Higashitani, K., {\it J. Chem. Eng. Japan\/}
{\bf 12}, 460 (1979).

\NI\hang 33. Ludwig, F.-P. and Schmelzer,  J.,
{\it J. Colloid Interface Sci.\/} {\bf 181}, 503 (1996).

\NI\hang 34. Overbeek, J. Th. G.,
{\it Adv. Colloid Interface Sci.\/} {\bf 15}, 251 (1982).

\NI\hang 35. Krug, J. and Spohn, H., {\it in\/} ``Solids Far from 
Equilibrium'' (C. Godrech{\` e}, Ed.), Cambridge University Press, 1991.

\NI\hang 36. Family, F. and Vicsek, T., ``Dynamics of Fractal 
Surfaces,'' World Scientific, Singapore, 1991.

\NI\hang 37. Medina, E., Hwa, T., Kardar, M. and Zhang, Y.-C.,
{\it Phys. Rev. A\/} {\bf 39}, 3053 (1989).

\NI\hang 38. Ruckenstein, E. and Prieve, D. C., {\it in\/} ``Testing 
and Characterization of Powders and Fine Particles,'' (J. K. Beddow 
and T. P. Meloy, Eds.), Heyden, London, 1980.

\NI\hang 39. Binder, K., {\it in\/} ``Alloy Phase Stability,'' (G. M. Stocks and A. Gonis, Eds.), Kluwer Academic Publishers, Boston, 1989.

\NI\hang 40. Ball, R. C. and Jullien, R.,
{\it J. Physique Lett.\/} {\bf 45}, 103 (1984).

\NI\hang 41. Sander, L. M., {\it in\/} ``Multiple Scattering of Waves 
in Random Media and Random Rough Surfaces,'' p. 51, The Pennsylvania 
State University publication, 1985.

\NI\hang 42. Stauffer, D. and Aharony, A., ``Introduction to Percolation Theory,'' Springer, Berlin, 1992.

\NI\hang 43. Jullien, R. and Meakin, P.,
{\it Europhys. Lett.\/} {\bf 4}, 1385 (1987).

\NI\hang 44. Privman, V. and Barma, M.,
{\it J. Chem. Phys.\/} {\bf 97}, 6714 (1992).

\NI\hang 45. Privman, V. and Nielaba, P.,
{\it Europhys. Lett.\/} {\bf 18}, 673 (1992).

\NI\hang 46. Nielaba, P. and Privman, V., 
{\it Modern Phys. Lett. B\/} {\bf 6}, 533 (1992).

\NI\hang 47. Evans, J. W.,
{\it Vacuum\/} {\bf 41}, 479 (1990).

\NI\hang 48. Xiao, R.-F., Alexander, J. I. D. and Rosenberger, F.,
{\it Phys. Rev. A\/} {\bf 45}, R571 (1992).

\NI\hang 49. Weiss,  G. H.,
{\it J. Statist. Phys.\/} {\bf 42}, 3 (1986).

\NI\hang 50. German, R. M., ``Particle Packing Characteristics,'' 
Metal Powder Industries Federation, Princeton, 1989.

\NI\hang 51. Brilliantov, N. V. and Krapivsky,  P. L.,
{\it J. Phys. A\/} {\bf 24}, 4787 (1991).

\NI\hang 52. Schaefer, D. W., Martin, J. E., Wiltzius, P. and
Cannell, D. S., {\it Phys. Rev. Lett.\/} {\bf 52}, 2371 (1984).

\NI\hang 53. Kelton, K. F., Greer, A. L. and Thompson, C. V.,
{\it J. Chem. Phys.\/} {\bf 79}, 6261 (1983).

\NI\hang 54. Reif, F., ``Fundamentals of Statistical and Thermal 
Physics,'' McGraw-Hill, New York, 1965.

\NI\hang 55. Goia, D. V. and Matijevi{\' c}, E., 
{\it Colloids Surf.\/} (submitted, 1998).

\NI\hang 56. Turkevich, J., Stevenson, P. C. and Hillier, J.,
{\it Discuss. Faraday Soc.\/} {\bf 11}, 55 (1951).

\NI\hang 57. Keim, R., Ed., ``Gmelin Handbook of Inorganic and 
Organometallic Chemistry, Gold,'' 8$^{\rm th}$ ed., Suppl. 
Vol. B1, p. 235-278, Springer-Verlag, Berlin, 1992.

\NI\hang 58. Weiser, H. B., ``Inorganic Colloid Chemistry,'' 
Vol. I, Ch. II, Wiley, New York, 1933.

\NI\hang 59. Linke, W. F., ``Solubilities of Inorganic and Metal-Organic
Compounds,'' 4$^{\rm th}$ ed., Vol. 1, p. 243, Van Hostrand, 
Princeton, 1958.

\NI\hang 60. Cotton, F. A., Wilkinson, G. and Gauss, P. L.,
``Basic Inorganic Chemistry,'' 3$^{\rm rd}$ ed., p. 61, Wiley, 
New York, 1995.

\NI\hang 61. ``Gmelins Handbuch der Anorganischen 
Chemie,'' 8$^{\rm th}$ ed., p. 429, Verlag Chemie, Weinheim, 1954.

\NI\hang 62. Gray, D. E., Ed., ``American Institute of Physics 
Handbook,'' 3$^{\rm rd}$ ed., p. 2-208, 1957.

}

\NP

\centerline{\bf Figure Captions}

\NI\hang Figure 1:\ \ \ Scanning electron micrographs, at two different magnifications, of final (secondary) gold particles
obtained by adding rapidly 100$\,$cm$^3$ of an aqueous solution of
HAuCl$_4$ (0.5$\,$mol$\,$dm$^{-3}$) into 400$\,$cm$^3$ of an aqueous
solution of ascorbic acid (0.5$\,$mol$\,$dm$^{-3}$), at room temperature; see (50).

\NI\hang Figure 2:\ \ \ (A) Field emission microscopy image of gold particles shown in Figure~1. (B) Enlarged image of the darkened area in (A).

\NI\hang Figure 3:\ \ \ Particle size distribution of primary particles (singlets) of gold prepared in a dilute system ($2 \cdot 10^{-4}\,$mol$\,$dm$^{-3}$ HAuCl$_4$), determined by dynamic light scattering.

\NI\hang Figure 4:\ \ \ Distribution of the secondary particles by their sizes, calculated using $\sigma=0.51\,$N$/$m,
for times $t=0.001$, 0.01, 0.1, 1, 10 sec. Parameter values used in the numerical calculation
are given in Section 5.

\NI\hang Figure 5:\ \ \ The same plot as in Figure~4, using $\sigma=0.57\,$N$/$m,
for times $t=0.1$, 1, 10 sec.
  
\NI\hang Figure 6:\ \ \ The same plot as in Figure~4, using $\sigma=0.63\,$N$/$m,
for times $t=0.1$, 1, 10 sec.
  
\NI\hang Figure 7:\ \ \ The function $\tau (t)$, defined in Section 5, calculated for the values of $\sigma$ corresponding to Figures 4-6.

\bye